\overfullrule 0pt
\vsize=9truein
\hsize=6.5truein

\def \A {{\hbar \omega^3 \over 2\pi^2 c^3}}

\def \Is {$I_*$}
\def \It	{$I_{\tau}$}
\def \Idt	{$I_{\Delta\tau}$}
\def \FF {{\bf F}}
\def \F	{{\bf f}}
\def \E	{{\bf E}}
\def \B	{{\bf B}}

\def \a	{{\bf a}}
\def \p  {{\bf p}}
\def \g  {{\bf g}}

\def \v  {{\bf v}}
\def \w {\omega}
\def \l {\lambda}
\def \ew {\eta(\omega)}

\def \Nzp	{{\bf N}^{zp}}
\def \Fzp	{{\bf f}^{zp}}
\def \Ezp	{{\bf E}^{zp}}
\def \Bzp	{{\bf B}^{zp}}
\def \pzp	{{\bf p}^{zp}}
\def \gzp	{{\bf g}^{zp}}
\def \D {\Delta}
\def \csh {\cosh \left( {a\t \over c} \right)}

\def \tnh {\tanh \left( {a\t \over c} \right)}

\def \snh {\sinh \left( {a\t \over c} \right)}
\def \snx {\sinh \left( {2 a\t \over c} \right)}

\def \t {\tau}
\def \b {\beta}
\def \bt {\beta_{\tau}}
\def \gt {\gamma_{\tau}}
\def \< {\left\langle}
\def \> {\right\rangle}
\def \[ {\left[ }
\def \] {\right] }
\def \( {\left(}
\def \) {\right)}

\def \Sm { \sum_{\lambda=1}^2 }

\def \he {\hat{\epsilon}}
\def \hk {\hat{k}}

\def \ra {\rightarrow}

\def \sn {\smallskip\noindent}
\def \pt {$(c^2/a, 0, 0)$}
\def \Hzp {\sqrt{\hbar \omega \over 2 \pi^2}}

\centerline{\bf Inertia as reaction of the vacuum to accelerated motion}
\bigskip
\centerline{Alfonso Rueda}
\centerline{\it Department of Electrical Engineering \& Department of Physics, ECS
Building}
\centerline{\it California State University, 1250 Bellflower Blvd.,
Long Beach, California 90840}
\centerline{arueda@csulb.edu}
\bigskip
\centerline{Bernhard Haisch}
\centerline{\it Solar and Astrophysics Laboratory, Dept. H1-12, Bldg.
252,  Lockheed Martin}
\centerline{\it 3251 Hanover Street, Palo Alto, California 94304}
\centerline{and}
\centerline{\it Max-Planck-Institut f\"ur Extraterrestrische Physik,
D-85740 Garching, Germany}
\centerline{haisch@starspot.com}
\bigskip
\centerline{(Physics Letters A, in press)}
\bigskip
\centerline{\bf Abstract}
\bigskip
It was proposed by Haisch, Rueda and Puthoff that the inertia of matter could be
interpreted at least in part as a reaction force originating in interactions between
the electromagnetic zero-point field (ZPF) and the elementary
charged consitutents (quarks and electrons) of matter. Within the limited context
of that analysis, it appeared that Newton's equation of motion ({\bf f} = m{\bf a})
could be inferred from Maxwell's equations as applied to the ZPF, i.e. the
stochastic electrodynamics (SED) version of the quantum vacuum. We report
on a new approach which avoids the {\it ad hoc} particle-field interaction model
(Planck oscillator) of that analysis, as well as its concomitant
formulational complexity. Instead, it is shown that a non-zero ZPF momentum flux
arises naturally in accelerating coordinate frames from the standard relativistic
transformations of electromagnetic fields. Scattering of this ZPF momentum flux by
an object will yield a reaction force that may be interpreted as a contribution to
the object's inertia. This new formulation is properly covariant yielding the
relativistic equation of motion:
${\cal F}={d{\cal P} / d\t}$. Our approach is related by the principle of
equivalence to Sakharov's conjecture of a connection between Einstein action and
the vacuum. If correct, this concept would substitute for Mach's
principle and imply that no further mass-giving Higgs-type fields may be required to
explain the inertia of material objects, although extensions to include the
zero-point fields of the other fundamental interactions may be necessary for a
complete theory of inertia.


\vfill\eject
\noindent{\bf 1. Introduction}

\bigskip
Vigier [1] has recently presented a discussion of the ``unsolved mystery
in modern physics'' known as inertia, the instantaneous opposition to acceleration
of all material objects. In the view of Newton it was an inherent property of
matter for which no further explanation was possible. In the nineteenth century
Mach proposed, on the basis of relativity of all motion, that inertia somehow
originated in a collective linkage of all matter in the Universe. No
successful quantitative formulation of Mach's principle has ever been developed [2].
A preliminary attempt by Sciama [3] resulted in a prediction that was later shown
to be contradicted by observations. In his formulation, the asymmetrical
distribution of surrounding matter in the Milky Way galaxy should give rise to a
directional dependence of inertial mass of order $\Delta m/m = 10^{-7}$. This
prediction is contradicted by the experiments of Hughes and Drever showing that
$\Delta m/m \le  10^{-20}$ [4]. This experiment furthermore shows that the material
entity responsible for inertia, if other than the particle itself, displays a very
remarkable degree of isotropy.

\bigskip
It is well known that general relativity does not embody Mach's principle:
Solutions of the field equations are possible for an empty universe and for a
rotating universe. Additional conflicts with Mach's principle have been presented
by Vigier [1] and by Rindler [5].
An additional motivation for finding a basis of inertia involving
locally-originating forces may be preservation of causality, since Mach's principle
would appear to call for instantaneous action at a distance of some sort. The view
proposed herein, which is presented in full detail by the authors elsewhere
[6], proposes
to substitute for Mach's principle a local electrodynamic interaction which is
perfectly consistent with causality.

\bigskip
We concentrate solely on the electromagnetic vacuum, leaving the almost certain
contributions of other vacuum fields, such as the case of the Dirac vacuum discussed
by Vigier [1], for further extensions of the theory. The original development of
this concept by Haisch, Rueda and Puthoff [7] depended upon Lorentz force
interactions between the charged particles constituting matter (quarks and
electrons) and the electromagnetic ZPF. The more general formulation presented
herein is independent of specific particle-field interactions. All that is needed
to generate an acceleration-dependent reaction force that may be interpreted as
inertia is for a scattering-like process of the ZPF radiation to occur in any
material object undergoing acceleration. The inertia of composite particles such as
protons and neutrons would arise via ZPF scattering at the level of the individual
quarks. The neutrino, being apparently a truly neutral particle, should not have
any inertial mass, consistent with current expectations. The theory at this
early stage also does not address certain properties of bosons, such as the
gravitational deflection of photons, and the apparent masses of the $Z^0$,
$W^+$ and
$W^-$ bosons mediating the weak interaction.

\bigskip
The concept we present is a descendent of a conjecture by Sakharov for a
connection of Einstein action and the vacuum [8]. The principle of equivalence
would imply that gravitation and inertia must have a similar connection to the ZPF.
A preliminary development of the Sakharov conjecture in the context of SED was
carried out by Puthoff [9]. While Puthoff's concept we view as promising, certain
problems remain [9].  Nevertheless the general Sakharov and Puthoff concept does
answer the potential criticism that a real, inertia-generating ZPF should generate
an unacceptably large cosmological constant. This is not the case, since it would be
the effect of the ZPF on charged particles that would generate gravitation in the
Sakharov-Puthoff approach, not the ZPF in and of itself. The energy density of the
ZPF cannot be equivalent to gravitating mass; for further discussion of the
astrophysical implications see Haisch and Rueda [10].

\bigskip
\vfill\eject
\noindent{\bf 2. Inertia as a Reaction Force}

\bigskip
Newton's second law may be most generally written as

$$\F={d\p \over dt}=\lim_{\D t \ra 0} {\D \p \over \D t}  ,
\eqno(1)$$

\sn which is the limiting form of the space part of the relativistic
four-force:

$$\FF={d\p \over d\t}= \gamma{d\p \over dt} ,
\eqno(2)$$

\sn which for the case when $\beta \ra 0$ and $\gamma \ra 1$
becomes Eq. (1)

\bigskip
It is seen that the second law is a {\it definition} of force as
the rate of change of momentum imparted to an object by an agent. Having
defined force, Newton's third law states that such a force will result in the
creation of an equal and opposite reaction force back upon the accelerating agent. 
This now makes the concept of inertia a necessity: Inertia must be
attributed to the accelerating object in order to generate the equal and opposite
reaction force upon the agent required by the third law.  It is our proposition
that resistance from the vacuum is the physical basis of that reaction force. 
One can interpret this as either the origin of inertia of matter or as a
substitute for the concept of innate inertia of matter.  Inertia becomes
a placeholder for this heretofore undiscovered vacuum-based reaction force which is
a necessary requirement of Newton's third law.  Force is then seen to be a
primary concept; inertia is not.

\bigskip
This can be made explicit as follows.
Newton's third law is a statement about
symmetry in nature for contact forces such that an applied
force $\F$, must necessarily result in a reaction force $\F_r$

$$\F=-\F_r .
\eqno(3)$$

\sn Inertia as the dynamical extension of this law can be made
explicit by writing the $\F=m\a$ relation as

$$\F=-(-m\a) ,
\eqno(4)$$

\sn which makes it clear that inertia as a resistance to acceleration
is equivalent to a reaction force of the form

$$\F_r=-m\a .
\eqno(5)$$

\bigskip
It is an
experimental fact that to accelerate an object a force must be
applied by an agent and that the agent will thus experience an equal
and opposite force so long as the acceleration continues.  We argue
that this equal and opposite force also has a deeper physical cause: the scattering
of or interaction with ZPF radiation.  We demonstrate that from the point of view of
a  nearby inertial observer there exists a net energy and momentum flux
(derived from the Poynting vector) of ZPF radiation transiting the accelerating
object in a direction necessarily opposite to the acceleration vector.  The
scattering opacity of the object to the transiting flux creates the
back-reaction force customarily called the inertial reaction.  Inertia is
thus, in part, a special kind of electromagnetic drag effect, namely one that
is acceleration-dependent since only in accelerating frames is the
ZPF perceived as asymmetric.   In stationary or uniform-motion frames
the ZPF is perfectly isotropic.

\bigskip
Following the approach in [7] and [11] we consider the case of uniform acceleration,
{\bf a}, which results in hyperbolic motion [12]. We define three coordinate
systems. $S$ is a non-inertial frame in which a uniformly accelerating object is
fixed at the point \pt. \Is \ is the inertial laboratory frame. \It \ is a series
of inertial frames whose  \pt \ point at proper time $\t$ corresponds
instantaneously with object point \pt \ in $S$, i.e. it is an instantaneously
co-moving frame. The acceleration of this point of
$S$ with respect to
\It \ is always {\bf a} for all proper times $\t$. At $\t=0$ the \pt \
point of $S$ also coincides with the
\pt \ point of \Is. The acceleration of the \pt \ point of $S$ as seen from \Is \ is
${\bf a}_*=\gt^{-3} {\bf a}$ [12].
We take the acceleration to be along the $x$-axis, ${\bf
a}=a\hat{x}$. The Rindler non-inertial frame $S$ is rigid and it 
has approximately the same constant acceleration within a small
neighborhood of the center of the accelerating object. (But as is well known the
acceleration is not by any means the same everywhere throughout $S$.)

\bigskip
With these coordinate definitions in place we can examine the object vs. ZPF
momentum relations that need to be satisfied to justify the proposition that an
electromagnetic reaction force can account for inertia. At proper
time $\D\t$ let an  object be instantaneously at rest in the inertial
coordinate frame \Idt \ at the point \pt \ of that frame. 
Moreover at the object proper time $\t = 0$ (that corresponds to the
time $t_* = 0$ of \Is), the object was instantaneously at
rest at the point \pt \ of the laboratory inertial frame.  After a short lapse of
laboratory time $\Delta t_*  > 0$ that corresponds to the object
proper time $\D\t$, the object is seen, from the viewpoint of \Is, to
have received from the accelerating agent the amount of impulse or
momentum increment $\D\p_*$.  The expression (1) but as seen in \Is
\ is thus

$$\F_*={d\p_* \over dt_*} = \lim_{\D t_* \ra 0} {\D \p_* \over \D
t_*} .
\eqno(6)$$

\sn
At the corresponding object proper time $\D\t$, the object is
instantaneously at rest in the comoving inertial frame \Idt.
Consequently the momentum of the object at proper time $\D\t$ as
viewed in \Idt \ is of course zero.

\bigskip
Our goal is to show that a ZPF electromagnetic reaction force will prove to be
the exact opposite of this, and can therefore reasonably be interpreted as the
inertia of the object, i.e. that in general

$$\Fzp=\F_r=-\F .
\eqno(7)$$

\sn
We will arrive at this by specifically considering the condition in the \Is \ frame,

$$\Fzp_*=\F_{r*}=-\F_* .
\eqno(8)$$

\sn The key is to find whether $\Fzp$ (or $\Fzp_*$) will prove, from relativistic
electrodynamics, to be proportional to  $-\a$. When we compare Eqs. (1), (5) and
(7), it follows that if the accelerating agent by means of the force $\F$ gives to
the object during object proper time interval $\D\t$ an impulse or change of
momentum
$\D\p$, there must be a corresponding impulse (change of momentum)
$\D\pzp$ provided by the ZPF in the opposite
direction to $\D\p$ so that 

$$\D\pzp \ = -\D\p 
\eqno(9)$$

\sn if our proposition is to be true.
Hence $\D\pzp$ is the matching reactive
counter-impulse given by the ZPF that opposes the impulse $\D\p$ given by the
accelerating agent.  We refer both $\D \pzp$ and $\D \p$ to the same inertial
frame and in this case to the laboratory frame \Is \ and  write as the required
condition,

$$\D \pzp_* \ = -\D \p_* .
\eqno(10).$$

\sn As this momentum change for the object $\D \p_*$ {\it is
calculated with respect to the inertial frame (that conventionally we
call the laboratory frame) \Is \ and not with respect to any other
frame}, (e.g., the  inertial frame \Idt) it is necessary to calculate
the putative ZPF-induced opposing impulse $\D \pzp_*$ with respect to
the {\it same} inertial frame \Is \ (and not with respect to \Idt \
or any other frame). We write

$$\D \pzp_*=\pzp_*(\D t_*)-\pzp_*(0)=\pzp_*(\D t_*) .
\eqno(11)$$

\sn The momentum $\pzp_*(\D t_*)$ is essentially the integral of
$d\pzp_*$ from \Is-frame time $t_* = 0$ to \Is-frame time $t_* =\D
t_*$.  The last equality follows  from symmetry of the ZPF
distribution as viewed in \Is \ that leads to

$$\pzp_*(0)=0 .
\eqno(12)$$

In what follows we seek  to find a mathematical expression for  the
ZPF-induced inertia reaction force $\Fzp_*$. For this purpose it is
useful to state that from Newton's third law and the force defined
above we can write that the following must be true if our hypothesis
is correct:

$$\lim_{\D t_* \ra 0} {\D \pzp_* \over \D t_*}=
\Fzp_*
 \ = -\F_* =-\lim_{\D t_* \ra 0} {\D
\p_* \over \D t_*} .
\eqno(13)$$

\sn If  the inertia origin propounded here is correct then Eq. (13),
at least in the subrelativistic case, should yield a nonvanishing
force $\Fzp_*$  that is parallel to the direction of the acceleration
$\a = a\hat{x}$, opposite to it, and proportional to the acceleration
magnitude $a = |\a|$.

\bigskip
\noindent{\bf 3. The ZPF in the Accelerating Frame}

\bigskip
We concern
ourselves  with the ZPF momentum flux entering an accelerating
object. Consider the
following simple fluid analogy involving as a heuristic device a
constant velocity and a spatially varying density (in place of the
usual hyperbolic motion through a uniform vacuum medium). Let a small
geometric figure of a fixed proper volume $V_0$ move uniformly with constant
subrelativistic velocity ${\bf v}$ along the $x$-direction.  The
volume $V_0$ we imagine as always immersed in a fluid that is
isotropic, homogeneous and at rest, except such that its density
$\rho(x)$ increases in the $x$-direction but is uniform in the $y$-
and $z$-directions.  Hence, as this small fixed volume $V_0$ moves in
the $x$-direction, the mass enclosed in its volume, $V_0 \rho(x)$,
increases.  In an inertial frame at rest with respect to the
geometric figure the mass of the volume, $V_0 \rho(x)$, is seen to
grow.  Concomitantly it is realized that the volume $V_0$ is sweeping
through the fluid and that this  $V_0 \rho(x)$ mass grows because
there is a net influx of mass coming into $V_0$ in a direction
opposite to the direction of the velocity.  In an analogous fashion,
for the more complex situation envisaged in this paper,
simultaneously with the steady growth of the ZPF momentum contained
within the volume of the object, the object is
sweeping through the ZPF of the \Is \ inertial observer and for him
there is a net influx of momentum density coming from the background
into the object and in a direction opposite to that of the velocity
of the object.

\bigskip
To calculate the ZPF momentum flux we transform the customary
SED representation of the ZPF from an inertial to an acclerating frame. For the
case of the hyperbolic motion [7][11][12], the velocity
$u_x(\t) = \bt c$ of the object point fixed in $S$ with respect to
\Is,  is

$$ \bt={u_x(\t) \over c} = \tnh  
\eqno(14)$$

\sn and then

$$ \gamma_{\t} = \left( 1-\beta_{\t}^2 \right)^{-1/2} = \csh .  
\eqno(15)$$

The ZPF in the laboratory system \Is \ is given by the standard SED
Fourier mode representation [7][11]

$$\Ezp ({\bf R}_*,t_*) = \Sm \int d^3 k \
\he({\bf k}, \l) \Hzp \cos [ {\bf k \cdot R}_* - \omega t_*
- \theta({\bf k},\l) ] ,
\eqno(16a)$$

$$\Bzp ({\bf R}_*,t_*) = \Sm \int d^3 k \ (\hk \times \he)
\Hzp  \cos [ {\bf k \cdot R}_* - \omega t_* - \theta({\bf
k},\l) ] .
\eqno(16b)$$

\sn
${\bf R}_*$ and $t_*$ refer
respectively to the space and time coordinates of the point of
observation of the  field  in  \Is. The phase term $\theta({\bf k},\l)$ is a
family of random variables, uniformly distributed between 0 and
2$\pi$, whose mutually independent elements are indexed by the
wavevector  ${\bf k}$ and the polarization index $\l$ (or more
technically, $\theta({\bf k},\l)$ is a stochastic process with index
set $\{ ({\bf k},\l) \}$).

\bigskip
A simple Lorentz rotation from \Is \ into \It \ allows us to calculate the $\Ezp$
and
$\Bzp$ in \It. We assume that the fields as seen in \It \ to also correspond to the
fields as {\it instantaneously} seen in $S$. A crucial point is the following.
Though the fields at the object
point in $S$ and in the corresponding point of the co-moving frame \It \
that instantaneously coincides with the object point are exactly the same, this
does not mean that detectors in
$S$ and in \It \ will be subject to the same effect, i.e., experience the
same radiation-field time evolution.  {\it Detectors need time to
perform their measurements}: This necessarily involves integration
over some interval of time and the evolution of the fields in $S$ and
in \It \ are obviously different.  Hence a detector at rest in \It \
and  the same detector at rest in $S$ do not experience the same
thing. Summarizing, while the two fields, namely that of $S$ and that of
\It, are the same at a given space-time point, the evolution of the
field in $S$ and the evolution of the field in \It \  are by no means
the same.  Furthermore any field or radiation measurements in \It \
and in $S$ both take some time and are not confined to a single
space-time point.

\bigskip
We clarify the notation used in the sense that all
polarization components are understood to be scalars, i.e.,
directional cosines, but written in the form
$\he_i ({\bf k},\l) \equiv \he \cdot \hat{x}_i$, where 
$\hat{x}_i=\hat{x},\hat{y},\hat{z}; \ i=x,y,z$, stands for three unit vectors along
the three space directions. The karat in $\he_i ({\bf k},\l)$ means that the
directional cosines come from axial projections of the polarization unit vector
$\he$ .  We use the same convention  for components of the $\hk$ unit
vector where, e.g., $\hk_x$ denotes $\hk \cdot \hat{x}$.  We can
select space and time coordinates and orientation in \Is \ such that
[7][11]

$${\bf R}_*(\t)\cdot \hat{x} = {c^2 \over a} \csh 
\eqno(17)$$

$$t_*={c \over a} \snh
\eqno(18)$$

After Lorentz-transforming the fields from \Is \ in Eq. (16) to those in \It \ and
using Eqs. (14), (15), (17) and (18) we obtain [6]

$$
\Ezp (0,\t) = \Sm \int d^3 k \times
$$
$$
\bigg\{
\hat{x}\he_x +
\hat{y}\csh  \[ \he_y - \tnh (\hk\times \he)_z \] +
\hat{z}\csh  \[ \he_z + \tnh (\hk\times \he)_y \]
\bigg\}
$$
$$
\times \Hzp \cos \[ k_x{c^2 \over a} \csh - {\w c \over a}
\ \snh - \theta({\bf k},\l) \]
\eqno(19a)$$

$$
\Bzp (0,\t) = \Sm \int d^3 k \times
$$
$$
\bigg\{
\hat{x}(\hk \times \he)_x +
\hat{y}\csh  \[ (\hk\times \he)_y + \tnh \he_z \] +
\hat{z}\csh  \[ (\hk\times \he)_z - \tnh \he_y \]
\bigg\}
$$
$$
\times \Hzp \cos \[ k_x{c^2 \over a} \csh - {\w c \over a}
\ \snh - \theta({\bf k},\l) \] .
\eqno(19b)$$

\sn This is the ZPF as instantaneously viewed from the object fixed
to the point $(c^2/a, 0, 0)$ of $S$ that is performing the hyperbolic
motion.

\bigskip
As it is the ZPF radiation background of \Is \ in the act of
being swept through by the object which we are calculating now, we
fix our attention on a fixed point of \Is, say the point of the
observer at \pt \ of \Is, that momentarily coincides with
the object at the object proper time $\t= 0$, and consider that point
as referred to the inertial frame \It \ that instantaneously will
coincide with the object at a future generalized  object  proper time
$\t>0$.  Hence  we  compute the \It-frame Poynting vector, but
evaluated at the \pt \ space point of the \Is \ inertial
frame, namely in \It \ at the  \It \ space-time point:

$$ct_{\t}={c^2 \over a}\snh ,
\eqno(20)$$
$$ x_{\t}=-{c^2 \over a}\csh,
\qquad y_{\t}=0,
\qquad z_{\t}=0.
\eqno(21)$$

\sn This Poynting vector we shall denote by $\Nzp_*$. Everything
however is ultimately referred to the \Is \ inertial frame as that is
the frame of the observer that looks at the object and whose ZPF
background the moving object is sweeping through.  In  order to
accomplish this we first compute

$$\eqalignno{
\< \Ezp_{\t}(0,\t) \times \Bzp_{\t}(0,\t) \> _x &=
 \< E_{y\t}B_{z\t}-E_{z\t}B_{y\t} \> \cr &= \gt^2 \< (E_{y*}-\bt
B_{z*})(B_{z*}-\bt E_{y*}) -(E_{z*}+\bt B_{y*})(B_{y*}+\bt E_{z*}) \>
\cr &=-\gt^2 \bt \< E_{y*}^2+B_{z*}^2+E_{z*}^2+B_{y*}^2 \>  +
\gt^2(1+\bt^2) \< E_{y*}B_{z*}-E_{z*}B_{y*} \> \cr &=-\gt^2 \bt \<
E_{y*}^2+B_{z*}^2+E_{z*}^2+B_{y*}^2 \> & (22)} $$

\sn that we use in the evaluation of the Poynting vector [6]

$$\Nzp_*={c \over 4\pi} <  \Ezp_{\t} \times \Bzp_{\t} >_* = \hat{x}
{c \over 4\pi} <  \Ezp_{\t} (0,\t) \times \Bzp_{\t} (0,\t) >_x .
\eqno(23)$$

\sn The integrals are now taken with respect to the \Is \  ZPF
background as that is the background that the \Is -observer
considers the object to be sweeping through.  This is why we denote
this Poynting vector as $\Nzp_*$, with an asterisk subindex instead
of a $\t$ subindex, to indicate that it refers to the ZPF of \Is. 
Observe that in the last equality of Eq. (22) the term proportional
to the $x$-projection of the ordinary ZPF Poynting vector of \Is \
vanishes.  The net amount of momentum of the background  the  object 
has  swept  through  after  a  time  $t_*$,  as  judged  again  from 
the  \Is-frame viewpoint, is

$$\pzp_* = \gzp_* V_* = {\Nzp_*  \over  c^2} V_* = -\hat{x}{1 \over
c^2}{c \over 4 \pi} \gt^2 \bt  {2 \over 3} \<  \E_*^2 + \B_*^2  \>
V_* .
\eqno(24)$$

\sn By means of Eq. (13) we will calculate the
force $\Fzp_*$ directly from the expression for $\pzp_*$.

\bigskip
\noindent{\bf 4. Momentum Flux and Newtonian Inertia}

\bigskip
Any observer at rest in an inertial frame sees the ZPF
isotropically distributed and thus the Poynting vector $\Nzp$
and the momentum density  $\gzp=\Nzp/c^2$ vanish.  This is of course the case for
the observer at rest in
\Is.  Consider now another inertial observer located at a geometric
point that, with respect to  \Is, moves uniformly with {\it constant}
velocity,  $\v= \hat{x} v_x = \hat{x} \b c$.  Imagine the instant of
time when the geometric point is passing and in the immediate
neighborhood of the stationary \Is \ observer.  Both observers
necessarily see the ZPF symmetrically and isotropically distributed
around themselves in their own frames.  However, the ZPF for each
observer is not, because of the Doppler shifts, isotropically
distributed with respect to the other frame.  The \Is-observer is located at the
center of his own
$k$-sphere, but the moving point is necessarily located off-center of
the \Is-observer's $k$-sphere [6]. Hence, for  the \Is-observer the ZPF
Poynting vector,  $\Nzp_*$, and the corresponding momentum density,
$\gzp_*$, impinging on the moving point should appear to be
non-vanishing.  Furthermore, because the motion of the geometric
point is uniform, not hyperbolic, both the $\Nzp_*$ and $\gzp_*$ at
the moving geometric point appear to the \Is-observer to be
time-independent constants of the motion. We interpret this as the basis of
the concept of momentum. A complete discussion is found in [6], particularly
Appendix B therein.

\bigskip Extend the consideration above to all the points inside
a small 
$\epsilon$-neighborhood of the previous geometric point that comove
with {\it constant} velocity $\v= \hat{x} c \b$.  Let $V_0$ be the
proper volume of that neighborhood.  Because of length contraction
such neighborhood has, in \Is, the volume $V_* = V_0/\gamma$. 
Clearly to the observer in \Is \ the neighborhood's $\gzp_*$ and 
$\Nzp_*$ do not appear as vanishing because of the uniform motion
with constant velocity,  $\v= \hat{x}\b c$, inducing Doppler shifts
of all the neighborhood's points with respect to \Is.   If the said
neighborhood exactly coincides with the location and geometry of  a 
moving  object of  proper volume  $V_0$  and  rest  mass $m _0$ that
has  the neighborhood's central geometric point at its center, then
according to ordinary mechanics, the object appears to the observer
in \Is \ as carrying a mechanical momentum  $\p_*=\gamma m_0 \v$.

\bigskip We turn now to the object's corresponding ZPF momentum. 
Because the object occupies its proper volume $V_0$ and coincides
with the uniformly moving 
$\epsilon$-neighborhood, it has for the observer at rest in \Is \ an
amount of ZPF momentum, $V_* \g_*=(V_0/\gamma)\g_*$, as described
above.  We re-emphasize that when measured and from the point of view
of the {\it inertial observer comoving with the object}, both the
object momentum and the Poynting vector of the ZPF do exactly vanish,
the last because in $k$-space the object is at the center of
that observer's $k$-sphere [6].  In the present case of a
constant velocity and zero acceleration for the object, as opposed to
the general case we have been considering of accelerated hyperbolic
motion, the momenta $\p_*$ and $\pzp_*$ above  are both of course
constants. Hence their time derivatives in Eq. (13) both vanish.

\bigskip
We
return to our original hyperbolic motion problem and  compute the Poynting
vector (a more complete discussion of this is found in Appendix A of [6])
that the  radiation should  have  at  the 
\pt \  point  of  \Is \  but referred to \It \ with the coordinates of Eq.
(21), viz,

$$\eqalignno{
\Nzp_*(\t) &={c \over 4\pi} \< \Ezp \times \Bzp \> \cr &=\hat{x}{c
\over 4\pi} \< E_y B_z - E_z B_y \>  \cr &=-\hat{x}{c \over
4\pi}{8\pi \over 3}\snx \int \A d\w &(25)}$$

\sn where $\Ezp$ and $\Bzp$ stand for $\Ezp_{\t}(0,\t)$  and
$\Bzp_{\t}(0,\t)$ respectively as in the case of Eq. (23)  and where
as in Eqs. (22), (23) and (24) the integration is understood to
proceed over the $k$-sphere of \Is.  The object now is not in uniform
but instead in accelerated motion.  If suddenly at proper time $\t$
the motion were to switch from hyperbolic back to uniform because the
accelerating action disappeared, we would  just need to replace in
Eq. (25) the constant rapidity $s$ at that instant for $a\t$, and
$\bt$ in Eq. (14) would then become $\tanh(s/c)$.  (But then $\Nzp$
would cease to be, for all times onward, a function of $\t$ and
force  expressions as Eq. (28) below would vanish.)  Observe that we
make explicit the $\t$ dependence of this as well as of  the
subsequent quantities below.  $\Nzp_*(\t)$ represents energy flux,
i.e., energy per unit area and per unit time in the $x$-direction. 
It also implies a parallel, $x$-directed momentum density, i.e.,
field momentum per unit volume incoming towards the object position,
\pt \ of $S$, at object proper time $\t$ and as estimated
from the viewpoint of \Is.  Explicitly such momentum density is 

$$\gzp_*(\t)={\Nzp_*(\t) \over c^2} = -\hat{x} {8\pi \over 3}{1 \over
4\pi c} \snx \int \ew \A d\w ,
\eqno(26)$$

\sn where we now introduce the henceforth frequency-dependent
coupling coefficient, $0 \le \ew \le 1$, that quantifies the fraction
of  absorption or scattering at each frequency.  Let $V_0$ be the
proper volume of the object, namely the volume that the object has in
the reference frame \It \ where it is instantaneously at rest at
proper time $\t$.  From the viewpoint of \Is, however, such volume is
then 
$V_*=V_0/\gt$ because of Lorentz contraction.  The amount of momentum
due to the radiation inside the volume of the object according to
\Is, i.e., the radiation momentum in the volume of the object viewed
at the laboratory is

$$\pzp_*(\t)=V_* \gzp_*={V_0 \over \gt} \gzp_*(\t)
 = -\hat{x}{4V_0 \over 3} c \bt \gt \left[ {1 \over c^2} \int \ew \A
d\w \right] , \eqno(27).$$

\sn which is again Eq. (24).

\bigskip At proper time $\t= 0$, the \pt \ point of the
laboratory inertial system \Is \ instantaneously coincides and
comoves with the object point of the Rindler frame $S$ in which the
object is fixed.  The observer located at $x_* = c^2/a, \ y_* = 0, \
z_* = 0$ instantaneously, at $t_* =0$, coincides and comoves with the
object but because the latter is accelerated with constant
acceleration $\a$, the object {\it according to} \Is \ should receive
a time rate of change of incoming ZPF momentum of the form:

$${d\pzp_* \over dt_*} = {1 \over \gt}{d\pzp_* \over d\t} \bigg|
_{\t=0} .
\eqno(28)$$

We {\it postulate} that such rate of change may be identified with a
force from the ZPF on the object.  Such interpretation, intuitively
at least, looks extremely natural. In this respect Rindler [12] in
introducing Newton's second law makes the following important
epistemological point: ``This is only `half' a law; for it is a mere
definition of force,'' and this is precisely the sense in which we
introduce it here as a definition of the force of reaction by the
ZPF.  If the object has a proper volume $V_0$, the force exerted on
the object by the radiation from the ZPF as seen in \Is \ at $t_* =0$
is then

$${d\pzp_* \over dt_*}=\Fzp_* =-\left[ {4 \over 3}{V_0 \over c^2}
\int \ew \A d\w \right] \a .
\eqno(29)$$

\sn Furthermore

$$m_i= \left[ {V_0 \over c^2} \int \ew \A d\w \right] 
\eqno(30)$$

\sn is an invariant scalar with the dimension of mass.  The expression for $m_i$
differs considerably from the corresponding one in [7] because here, on purpose,
no interaction features were included in the analysis. Such ZPF-particle
interactions will be taken up in future work. Observe that
in Eq. (30) we have neglected a factor of 4/3.  Such factor must be
neglected because a fully covariant analysis shows
that it disappears [6].  The corresponding form of $m_i$ as written (and
without the 4/3 factor) is then susceptible of a very natural
interpretation: Inertial mass of an object is that fraction of the 
energy of the ZPF radiation enclosed within the object that
interacts with it (parametrized by the $\ew$ factor in the integrand).

\bigskip
Clearly if
the acceleration  suddenly ceases at proper time $\t$, Eqs. (28) and (29)
identically vanish, signaling the fact that acceleration is the reason
that the vacuum produces the opposition that we identify with the
force of reaction known as inertia.  From the proper time instant
$\t$ when the acceleration $\a$ is turned off, the object continues
in uniform motion.  The object proceeds onwards with the rapidity $s$
it acquired up to that point, namely $a\t$.  Thus $\bt$ in Eq. (14)
and all quantities from Eqs. (25) to (27) become constants, as the
rapidity $s$ ceases to depend on the proper time $\t$.  Because of
the Lorentz invariance of the ZPF energy density spectrum [13], the
object is left at rest in the inertial frame \It \ and at the center
of the $k$-sphere of the \It \ observer but off-center of the
$k$-sphere of the \Is \ one [6]. From the \Is \ perspective
the object appears to possess a momentum (which reflects the ZPF
momentum inside $V_0$).  Observe furthermore that in Eq. (30) and previous equations
some cut-off procedure is implicit in that $\ew$ subsides at high
frequencies.

\bigskip
\noindent{\bf 5. Relativistic Force Expression}

\bigskip
The coefficient $m_i$ that we identify with the ZPF contribution
to inertial mass, corresponds then just to the ZPF-induced part of the rest mass of
the object. If the vacuum exerts an
opposition force on the accelerated object of magnitude
$-m_i\a$ as in Eq. (29) and if Newton's third law holds,
then the accelerating agent must exert an active  force $\F$  of
amount  $\F= m_i\a$ to produce the acceleration.  {\it This is the basis of Newton's
equation of motion.}  The radiative opposition made by the vacuum
precisely coincides time-wise with the onset of acceleration at every
point throughout the interior of the accelerated object, continues
exactly so long as the acceleration persists and is in direct
proportion to the amount of mass associated with that small region. 

\bigskip
It is important to add that our analysis
yields not just the nonrelativistic Newtonian case but it also
embodies a fully relativistic description within special relativity
[11] at least for the case of longitudinal forces, i.e., forces
parallel to the direction of motion.
Moreover the extension to the more general case where the accelerating or applied
force \F \ is non-uniform, (i.e., it changes both in magnitude and direction
throughout the motion of the object) is readily envisaged [6].

\bigskip
From the definition of the momentum $\pzp_*$ in Eq. (27),
from Eqs. (28), (29), and the force equation (8) it immediately
follows that the momentum of the object is

$$\p_*=m_i \gt \vec{\beta}_{\t} c ,
\eqno(31)$$

\sn in exact agreement with the momentum expression for a moving
object in special relativity.  The expression for the space
vector component of the four-force is then

$$\FF_*=\gt {d\p_* \over dt_*} = {d\p_* \over d\t} ,
\eqno(32)$$

\sn and as the force is pure in the sense of Rindler [11], the
correct form for the four-force immediately follows:

$${\cal F}={d{\cal P} \over d\t} = {d \over d\t}(\gt m_i c, \p) = \gt
\left( {1 \over c}{dE \over dt}, \F \right) = \gt \left( \F \cdot
\vec{ \beta}_{\t}, \F \right) = \left( \FF \cdot \vec{\beta}_{\t},
\FF \right) .
\eqno(33)$$

\sn
Consistency with Special
Relativity is established. (For a detailed exposition pertaining to
Eqs. 31--33 see [6].)

\bigskip
\noindent{\bf 6. Conclusions}

\bigskip

\bigskip
The new development here is simpler than that of [7]
in that it does not deal with
the dynamics of modeled particle-field interactions, but exclusively with the form
of the ZPF in relation to an accelerated object.  The final result is derived using
standard relativistic field transformations and does not involve
approximations. We extend the approch of [7] in deriving not only ${\bf f}=m{\bf
a}$ from Maxwell's equations as applied to the ZPF, but a properly relativistic
equation of motion, ${\cal F}={d{\cal P} / d\t}$. The fully covariant analysis
is presented in [6].

\bigskip
The inertia of protons and
neutrons would arise via ZPF scattering at the level of the individual quarks.
As presently formulated, our theory would not account for any possible neutrino
mass. Nor does the theory at this stage
address certain properties of bosons, such as the gravitational deflection of
photons, and the apparent masses of the $Z^0$,
$W^+$ and
$W^-$ bosons mediating the weak interaction.
One can naturally conjecture that analogous reaction forces interpreted as
inertial mass would arise in a more general way with the zero-point fluctuations of
other fields (like those of the weak and of the strong interactions).  The general
idea is that rather than postulating an {\it ad hoc} mass-giving field on top of
all the other fields, to examine instead if inertia can be explained by means of
the already well-established (vacuum) fields of one form or another, as e.g. the
approach of Vigier [1].

\bigskip
We very
explicitly used the ordinary notion of what force is.  So we cannot
claim any direct explanation of that concept, not even a
clarification of what force means.  With respect to this classical
force concept what we believe we have done is the following. 
Newton's third law requires that the motive force defined in the
second law be counterbalanced by a reaction force.  This has
traditionally been satisfied implicitly by assuming the existence of
inertia of matter.  We propose to have found an explicit origin for
this reaction force, viz. the acceleration-dependent scattering of
ZPF radiation that the accelerated object is forced to move into. 
Our analysis presupposed electrodynamics and special relativity and
other aspects of ordinary classical theory: Electrodynamics and some
aspects of special relativity have been used in our developments
since we used SED (that besides Maxwell's equations also presupposes
the Lorentz force).  As far as radiation reaction is concerned we
merely suspect that it is somewhat connected with the developments
here but so far this is
only a suspicion.

\bigskip
Finally we make two disclaimers.
We have used the methodology of SED. Recent work by Ibison and
Haisch [13] has resolved an important discrepancy between SED
and quantum electrodynamics (QED). Nevertheless a quantum theory-based derivation of
this proposal for inertia is highly desireable. Second, we are not
prepared to face the issue of how and in what sense our development might possibly
affect or relate to general relativity (beyond what was briefly mentioned
concerning Sakharov's hypothesis and the principle of equivalence). 

\bigskip
\centerline{\bf Acknowledgements}

\bigskip
A. R. thankfully acknowledges
detailed and extensive correspondence
with Dr. D. C. Cole that was instrumental in clarifying or
developing various arguments in this article.
B.H. wishes to thank Prof. J. Tr\"umper and
the Max-Planck-Insititut f\"ur Extraterrestrische Physik for
hospitality during several stays. 
We acknowledge support of this work by
NASA contract NASW-5050.

\vfill\eject

{

\bigskip

\noindent{\bf References}

\parskip=0pt plus 2pt minus 1pt\leftskip=0.25in\parindent=-.25in 

[1] J.-P. Vigier, Foundations of Physics, {\bf 25}, No. 10, 1461 (1995).

[2] W. H. McCrea, Nature {\bf 230}, 95 (1971). See also an attempt at an alternative
approach by R. C. Jennison and A. J. Drinkwater, J. Phys. A {\bf 10}, 167 (1977); see also J.
Barbour, ``Einstein and Mach's Principle'' in {\it Studies in the
History of General Relativity}, J. Eisenstadt and A. J. Knox (eds.)
(Birkhauser, Boston, 1988), pp. 125--153.

[3] D. W. Sciama, Mon. Not. Roy. Astr. Soc. {\bf 113}, 34 (1953); see also G.
Cocconi, and E. Salpeter, Il Nuovo Cimento, {\bf 10}, 646, (1958).

[4] S. Weinberg, {\it Gravitation and Cosmology: Principles and
Applications of the General Theory  of Relativity} (Wiley, New York,
1972), pp. 86--88.

[5] W. Rindler, Phys. Lett. A {\bf 187}, 236 (1994). There was a reply to this paper
by H. Bondi and J. Samuel, Phys. Lett. A, {\bf 228}, 121 (1997).

[6] A. Rueda and B. Haisch, Foundations of Physics, in press (1998). Detailed
analysis and necessary (but lengthy) derivations omitted in the present letter are
to be found here.

[7] B. Haisch, A. Rueda and H. E. Puthoff, Phys. Rev. A {\bf 49}, 678
(1994).  We also refer to this paper for review points and references
on the subject of inertia.

[8] A. D. Sakharov, Sov. Phys. Dokl. {\bf 12}, 1040 (1968); Theor.
Math. Phys. {\bf 23}, 435 (1975).  See also C. W. Misner, K. S.
Thorne and  J. A. Wheeler, {\it Gravitation} (Freeman, San Francisco,
1973) pp. 417--428.  

[9] H. E. Puthoff, Phys. Rev. A {\bf 39}, 2333 (1989); see also S. Carlip,
Phys. Rev. A {\bf 47}, 3452 (1993) and H.  E. Puthoff, Phys. Rev. A
{\bf 47}, 3454 (1993).  A detailed revision on the status of this last issue
has been carried out by D. C. Cole, K. Danley and A. Rueda (1998, in preparation)
and in a more limited context by K. Danley, M.S. Thesis, Cal. State Univ., Long
Beach (1994).  These works show that there remain unsettled questions in the
derivation of Newtonian gravitation.  However our inertia work and the equivalence
principle suggest to us that the vacuum approach to gravitation remains promising
once a more detailed relativistic particle model and analysis is implemented.

[10] B. Haisch \& A. Rueda, Astrophysical J., 488, 563 (1997).

[11] T. H. Boyer, Phys. Rev. D {\bf 29}, 1089 (1984); for clarity of
presentation the notation proposed in this article is followed here.

[12] W. Rindler, {\it Introduction to Special Relativity} (Oxford,
Clarendon 1991) pp. 91--93.  The most relevant part is Section 35,
pp. 90--93.  Hyperbolic motion is found in Section 14, pp. 33--36. 
Further details on hyperbolic motion are given in  F. Rohrlich, {\it
Classical Charged Particles} (Addison Wesley, Reading Mass, 1965) pp.
117 ff and 168 ff.  These are important references throughout this
paper.

[13]	The Lorentz invariance of the spectral energy density of the
classical electromagnetic ZPF  was independently  found by T. W.
Marshall, Proc. Camb. Phil. Soc. {\bf 61}, 537 (1965) and T. H.
Boyer, Phys. Rev. {\bf 182}, 1374 (1969); see also E. Santos, Nuovo
Cimento Lett. {\bf 4}, 497 (1972). From a quantum point of view every
Lorentz-invariant theory is expected to yield a Lorentz-invariant
vacuum.  The ZPF of QED should be expected to be Lorentz-invariant,
see, e.g., T. D. Lee, ``Is the physical vacuum a medium'' in {\it A
Festschrift for Maurice Goldhaber}, G. Feinberg,  A. W. Sunyar and
J. Wenesser (eds.), Trans. N.Y. Acad. Sci., Ser. II, Vol. 40 (1980).
For nice discussions on the Lorentz invariance of the ZPF and other
comments and references to related work in SED, see L. de la Pe\~na,
``Stochastic Electrodynamics: Its development, present situation and
perspective'' in {\it Stochastic Processes Applied to Physics and Other
Related Fields} (World Scientific, Singapore, 1983) B. Gomez et al
(editors) p. 428 ff. and also L. de la Pena and A. M. Cetto {\it The
Quantum Dice} (Kluwer, Dordrecht Holland, 1996) p. 113 ff. This last is the most
recent and comprehensive review on SED with some innovative features of its own
(for a review of this book see D. C. Cole and A. Rueda, Found. Phys. {\bf 26}, 1559,
1996).

[14] M. Ibison and B. Haisch, Phys. Rev. A {\bf 54},
2737 (1996).

}
\bye